\documentclass[aps,prl,twocolumn]{revtex4}
\usepackage{amsmath,bm,epsfig}

\def\Fbox#1{\vskip1ex\hbox to 8.5cm{\hfil\fboxsep0.3cm\fbox{%
  \parbox{8.0cm}{#1}}\hfil}\vskip1ex\noindent}  

\newcommand{\B}[1]{{\bm{#1}}}
\let \= \equiv \let\*\cdot \let\~\widetilde \let\^\widehat \let\-\overline

\def\<{\left\langle}    \def\>{\right\rangle}
\def\({\left(}          \def\){\right)}
 \def \[ {\left [} \def \] {\right ]}

\begin{document}
\title{Quantitative Theory of a Time-Correlation Function in a One-Component Glass-Forming Liquid with Anisotropic Potential}
\author{Edan Lerner, Itamar Procaccia and Ido Regev}
\affiliation{Department of Chemical Physics, The Weizmann
Institute of Science, Rehovot 76100, Israel }
\date{\today}
\begin{abstract}
The Shintani-Tanaka model is a glass-forming 
system whose constituents interact via anisotropic potential depending 
on the angle of a unit vector carried by each particle. The decay of 
time-correlation functions of the unit vectors exhibits the characteristics 
of generic relaxation functions during glass transitions. 
In particular it exhibits a 'stretched exponential' form,
with the stretching index $\beta$ 
{\bf depending strongly on the temperature}.
We construct a quantitative theory of this correlation function 
by analyzing all the physical processes that contribute to it, separating a rotational 
from a translational decay channel. 
Interestingly, the separate decay function of each of these processes is 
{\bf temperature independent}. Taken together with 
temperature-dependent weights 
determined {\em a-priori} by statistical mechanics one 
generates the observed correlation function in quantitative agreement with simulations 
at different temperatures. This underlines the danger of concluding anything about glassy 
relaxation functions without detailed physical scrutiny.
\end{abstract}
 \maketitle

The glass-transition is often described as 'mysterious' \cite{07Lan}; it is the opinion of the present authors
that there is nothing mysterious in the glass-transition, it is only necessary to penetrate the inhomogeneous 
states that develop naturally and understand the
role of a variety of dynamical processes that appear in various examples of glass transitions. Different systems 
may exhibit different ways of glassifying, although there are many generic aspects to identify the
phenomena as belonging to the same class. To underline this view we present in this Letter an in-detail
analysis of one very interesting model of glass-formation, exposing how a number of distinct process come together to exhibit a seemingly complex phenomenology. In particular we examine separately
the rotational and translational channel of relaxation, and how their relative contributions change
with decreasing the temperature. The Letter  culminates with a computation of the time correlation function as a sum of the simple processes with a-priori determined weights , in agreement with simulations at different temperatures. 
\begin{figure}
~\hskip -0.7 cm
\includegraphics[width=0.5\textwidth]{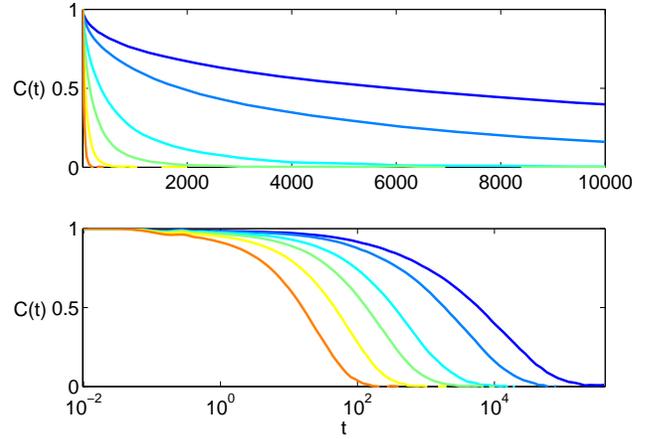}
\caption{(Color online). The relaxation function $C_R(t)$ as a function of linear time (upper panel), and of logarithmic time in the lower panel. The leftmost curve (in red) pertains to $T=0.3$ and in order to the right, the temperatures are $T= 0.25$, 0.22, 0.20, 018 and 0.17. Note the extreme slowing down in this range of temperatures.}
\label{data}
\end{figure}

{\bf The model} we employ here \cite{06ST} has particles of mass $m$ interacting via the potential
\begin{equation}
U(r_{ij}, \theta_i,\theta_j) = \overline{U}(r_{ij}) +\Delta U(r_{ij}, \theta_i,\theta_j) \ , \label{pot}
\end{equation}
where $r_{ij}$ is the distance between the two particles $i$ and $j$. The first term on the RHS of (\ref{pot}) is the standard isotropic
Lennard-Jones potential $\overline{U}_{ij} =4 \epsilon\left[ \left(\frac{\sigma}{r_{ij}}\right)^{12}- \left(\frac{\sigma}{r_{ij}}\right)^{6}\right]$, whereas the anisotropic part of the potential is given by
\begin{eqnarray}
&&\Delta U(r_{ij}, \theta_i,\theta_j)=-4\epsilon \Delta \left(\frac{\sigma}{r_{ij}}\right)^{6}\Big[h\left(\frac{\theta_i-\theta_0}{\theta_c}\right)\nonumber\\&&+h\left(\frac{\theta_j-\theta_0}{\theta_c}\right)-\frac{64}{35\pi}\theta_c\Big] \  , \label{anisopot}\\
&&h(x)=(1-x^2)^3~{\rm for}~|x|<1\ ; h(x) =0\,{\rm for}\,|x|\ge 1 \ . \nonumber
\end{eqnarray}
Here $\theta_i$ ($\theta_j$) is the included angle between the relative vector $\B r_{ij}\equiv \B r_i-\B r_j$ and a unit vector $\B u_i$ ($\B u_j$) (referred to below as `spin') which represents the orientation of the axis of particle $i$ ($j$). The function $h((\theta-\theta_0)/\theta_c)$ (with $\theta_0=126^{o}$ and $\theta_c=53.1^{o}$) has a maximum at $\theta=\theta_0$, and thus $\theta_0$ is a favored value of $\theta_i$. Thus the anisotropic term in the potential favors structures of five-fold symmetry. The parameter $\Delta$ controls the tendency of five-fold symmetry, and therefore of the frustration against crystallization. The units of mass, length, time and temperature are $m$, $\sigma$, $\tau=\sigma\sqrt{m/\epsilon}$ and $\epsilon/k_B$, respectively, with $k_B$ being Boltzmann's constant. According to the numerical simulations presented in \cite{06ST}, for $\Delta<0.6$ this system crystallizes upon reducing the temperature. The ground state crystal has an elongated hexagonal structure with
anti-ferromagnetic ordering of the spins $\B u_i$. For $\Delta\ge0.6$ the system fails to crystallize upon cooling, at least for the simulations times reported in \cite{06ST}. 

{\bf The relaxation function}: The time auto-correlation function of interest was introduced in \cite{06ST} in terms
of the spins, in the form 
\begin{equation}
C_R(t)\equiv (1/N)\sum_i \langle  \B u_i(t)\cdot \B u_i(0)\rangle \ . \label{corrfun}
\end{equation} 
Our own numerical results for this function are shown in Fig. \ref{data}.
As is customary in the field \cite{96EAN} the measured correlation function  was
fitted in \cite{06ST} to a stretched exponential form $C_R(t)\propto \exp[-(t/\tau_\alpha)^\beta]$. For $\Delta=0.6$ the relaxation is of Arrhenius form with a constant value $\beta\approx 0.95$ for $T>T_m=0.46$, but $\beta$ was fitted separately for every temperature  $T<T_m$ where it decreases with temperature. The relaxation times were fitted to a Vogel-Fulcher law $\tau_\alpha=\tau_0 \exp[DT_0/(T-T_0)]$; below we will show that we can reconstruct the correlation function from elementary processes with {\bf temperature-independent relaxation functions} (i.e. without needing to fit anything at different temperatures). As stated
previously \cite{08EP,08IPRS,08LP}, the Vogel-Fulcher fit is deeply misleading in indicating a finite temperature where the relaxation time diverge. We do not expect any singularity here for any dynamical process for any $T>0$. 

\begin{figure}
\hskip -0.4 cm
\epsfig{width=.45\textwidth,file=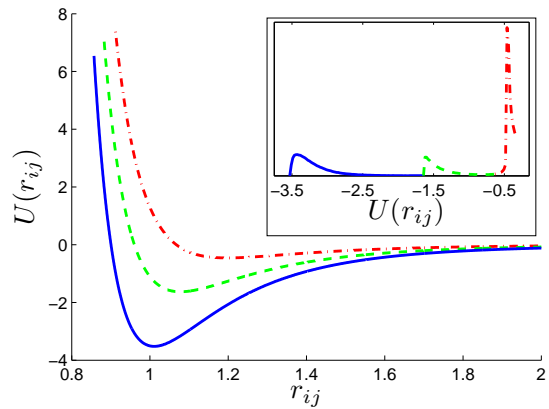}
\caption{(Color online). Potential curves for particle-pairs with two spins, one spin or no spin in favored
position (blue continuous line, green dashed or red dash-dotted line respectively). Inset: 
the measured energies of particle pairs, falling in three distinct ranges with gaps between them.}
\label{potentials}
\end{figure}

{\bf Statistical Mechanics}: To reconstruct the correlation function from elementary contributions we recall 
how statistical mechanics is constructed for this system \cite{07ILLP}.  In Fig. \ref{potentials}
we present the three potentials between two particles, depending on the orientation of their spins 
relative to the inter-particle vector distance: lowest in energy  (in blue continuous line) is 
the case for which both have a favored spin orientation; middle, in green dashed line (high, in red dash-dotted line)
is the potential when one (none) of the spins are in a favored orientation. One sees that the minima 
of these potentials occur with significant gaps in their energies, 
allowing us to now measure the {\em average} energy of {\em pairs} of particles as a function of temperature. These averages fall in three distinct ranges, such that the range of variation of each energy is much smaller than the gaps between the energies, see inset in Fig. \ref{potentials}. We denote the three effective energies
below as $2E_b$ $2E_g$ and $2E_r$ respectively. Next,  Ref \cite{07ILLP} defined quasi-species 
\begin{figure}
\hskip -1.25 cm
\epsfig{width=.30\textwidth,file=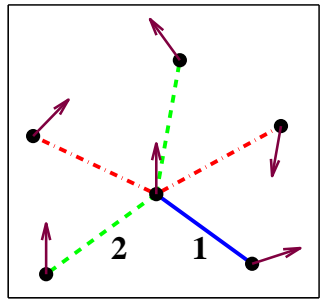}
\caption{(Color online). An example of an $n$-star with $n=5$,$i=2$, $j=2$ and $k=1$. The central particle has a spin with favored orientation with respect to edges 1 and 2. Thus these edges can be either blue or green, and this central spin cannot be favored with respect to any other edge. In the interesting range of 
temperatures we observe 36 $n$-stars with $4\le i+j+k\le 6$.}
\label{star}
\end{figure}
in the form of an $n$-stars, each of which is a given particle decorated by the $n$ inter-particle vector distances (edges) to its $n$ neighbors, see for example Fig. \ref{star}. Each such edge is colored according to the spin orientations. We denote by $i,j,k$ the number of red, green and blue edges such that $n=i+j+k$. It turns out that in the temperature
range of interest ($0<T<0.5$), in an overwhelming majority of $n$-stars (more than 98\%) the central particle has a spin orientation that is favorable with respect to two of its edges, leading to the
constraint (cf. \cite{08LP})
\begin{equation}
\sum_{ijk} (j+2k)c_{ijk}= 4 \ ,\label{constraint}
\end{equation}
which is important for the statistical mechanics of this system. The energy of 
an $n$-star (referred to as a quasi-particle) is computed as
\begin{equation}
E_{ijk} = iE_r +jE_g+kE_b \ , \label{energies}
\end{equation}
where $k\le 2$. Note that since the energies on the RHS of Eq. (\ref{energies}) depend on temperature, so does the energy of the quasi-particles. Notwithstanding, in the interesting temperature range the temperature dependence is weak;  we take the energies of the quasi-particles as $T$-independent. The degeneracy $g_{ijk}$ of the energy level (number of quasi-particles with the same energy) was computed in \cite{07ILLP} in the form
\begin{equation}
g_{ijk} = {2\choose k} {i+j+k-2\choose i} 2^{j+2k-2} 4^{i-k+2}\ .
\end{equation}

Finally we can write the partition function of the system:
\begin{equation}
Z(T,\lambda(T)) \equiv \sum_{ijk} g_{ijk} e^{-\beta E_{ijk}} e^{-\lambda(j+2k)} \ . \label{PF}
\end{equation}
The Lagrange multiplier $\lambda$ is introduced to insure that the constraint (\ref{constraint}) is satisfied. In terms of the partition function the mol-fraction of quasi-particles is 
\begin{equation}
c_{ijk} =\frac{g_{ijk} e^{-\beta E_{ijk}} e^{-\lambda(j+2k)}}{ Z(T,\lambda(T))} \ .  \label{mol-fractions}
\end{equation}
In \cite{08LP} it was shown that the prediction of this formula agrees well with simulations until
the system is 'jammed', or more correctly, until the simulation time is too short to allow the system to 
equilibrate. Here we will use this theory to compute the spin auto-correlation function. As noted in \cite{07ILLP},
we do not need to discuss separately all the 36 quasi-species, it suffices to bunch groups of $c_{ijk}$ together into groups  with $k=0,1,2$. This bunching is natural since it stresses the different environments
(potential barriers for unit vector orientation ) seen by the quasi-species in each group, helping us to
disentangle the competing dynamics leading to relaxation. 

{\bf Spin decorrealation}:  we first note that a spin can change its angle with respect to an interparticle vector distance either due to the spin rotation with the interparticle vector distance fixed, or due to translation in which the interparticle vector distance changes, cf.  \cite{96EAN} Fig. 6. We assert that to a good approximation these relaxation channels are independent and competing, allowing
us to write the relaxation function for each group  $k=0,1,2$ of quasi-particles in a product form:
\begin{equation}
F_k(\frac{t}{\tau_{\rm rot}(k,T)},\frac{t}{\tau_{\rm tr}(k,T)}) \approx f_{\rm rot} (\frac{t}{\tau_{\rm rot}(k,T)}) \times f_{\rm tr}(\frac{t}{\tau_{\rm tr}(k,T)}) \ ,
\end{equation}
where the subscripts 'rot'  and 'tr' stand for the rotational and translational decay channels respectively. We reiterate that these functions should be temperature independent except through the temperature dependence of the typical relaxation times $\tau_{\rm rot}(k,T)$ and $\tau_{\rm tr}(k,T)$.

Modeling the translation channel is the same for all the $k$ groups, since the slowing down
of translation in this model is dominated by the decrease in concentration of the $k=0$ quasi-species as the temperature decreases \cite{07ILLP}. Indeed, with the concentration of $c_{k=0}(T)$ decreasing, one defines a typical increasing scale $\xi(T)  \equiv 1/\sqrt{c_{k=0}(T)}$ whose physical meaning is the typical length of the cooperative
event that results in any appreciable translational motion. Accordingly the typical relaxation time associated with translation grows like \cite{08EP}
\begin{equation}
\tau_{\rm tr}(k,T) =\tau^0_{\rm tr}(k) e^{\frac{\mu \xi(T)}{T}}  \ ,
\end{equation}
where $\mu$ is the characteristic free energy per particle involved in the cooperative event of length $\xi$ and $\tau^0_{\rm tr}$ is
an attempt time (of the order of unity) that may depend on $k$. The relaxation function  $f_{\rm tr}(t/\tau_{\rm tr}(k))$
 is a simple exponential, 
 \begin{equation}
 f_{\rm tr}(t/\tau_{\rm tr}(k,T))=e^{-t/\tau_{\rm tr}(k,T)} \ ,  \label{ftr}
 \end{equation}
 and all the non-Arrhenius dependence comes from the dependence of $\xi$ on $T$.
 
 The rotation channel calls for more scrutiny, since the $k=0$ and $k=1$ quasi-species differ significantly
 from the $k=2$ quasi-species, as the latter tend to aggregate in clusters. We therefore need to deal with
 them differently.  Quasi-particles with $k=0,1$ are relatively free to rotate, and all that they need to do is
 to overcome the potential barrier for rotation. We thus expect their relaxation times to be of simple Arrhenius form, i.e.
\begin{equation}
\tau_{\rm rot}(k) = \tau^0_{\rm rot}(k)e^{\Delta E_{\rm rot}(k)/T} \quad \text{for }k=0,1\ .
\end{equation}
We expect $\Delta E_{\rm rot}(k)$ to be of the order of a bond energy in both cases, which is indeed what is found, cf. Table~\ref{tab1}.
The functional form of the relaxation functions turns out to be stretched exponentials, i.e.
\begin{equation}
 f_{\rm rot} (t/\tau_{\rm rot}(k,T)) \approx e^{-(t/\tau_{\rm rot}(k,T))^{\beta(k)}} \ , \quad\text{for }k=0,1 \ .
\end{equation}
The values of the temperature independent parameters are shown in Table \ref{tab1}.

\begin{figure}
\centering
\includegraphics[width=0.45\textwidth]{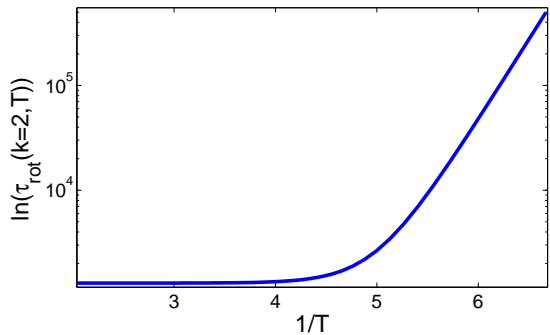}
\caption{(Color online). The logarithm of the relaxation time $\tau_{\rm rot}(k=2)$ as a function of 
inverse temperature}
\label{logt1}
\end{figure}

For $k=2$ we cannot expect such a simple model to hold. The reason is that at lower temperatures the $k=2$ quasi-particles aggregate inside clusters whose average size increases when the temperature decreases  \cite{06ST}. The distribution of cluster sizes depends on $T$, and to represent the rotational relaxation function one needs to decompose it into cluster contributions \cite{08HP,08LP}. Each cluster may decay simply with Arrhenius decay time, but the ensemble is expected to show a strongly non-Arrhenius relaxation time, as shown in \cite{08HP,08LP}. For the sake of brevity we do not  attempt here to derive the rotational relaxation function of the $k=2$ quasi-particles but we simply measure it to find
\begin{equation}
 f_{\rm rot} (t/\tau_{\rm rot}(k=2,T)) \approx \exp[-(t/\tau_{\rm rot}(k=2,T))^{0.6}] \ , \label{f2rot}
\end{equation}
with the measured value of $\tau_{\rm rot}(k=2,T)$ as shown in Fig. \ref{logt1}.
Note that this relaxation function is again temperature-independent (except through the dependence of $\tau_{\rm rot}(k=2)$)

We thus posses models for all the elementary relaxation functions in addition to Eq. (\ref{f2rot}). The parameter used are temperature independent, and are summarized in table \ref{tab1}. All that remains is to sum up the contributions
together with the right weights.
\begin{table}[!h]
\centering
\caption{Parameters used in the model}
\begin{tabular}{|c|c|c|c|}
\hline
 &$k=0$&$k=1$&$k=2$ \\
\hline
$\beta$&0.57 & 0.45 & 0.6 \\
\hline
$\Delta E_{\rm rot}(k)$&1.27 & 1.33 & - \\
\hline
$\mu$& 0.27 & 0.32 & 0.39 \\
\hline
$\tau^0_{\rm rot}(k)$&0.24 &2.48 & - \\
\hline
$\tau^0_{\rm tr}(k)$&2.48 & 1.6 & 1.07 \\
\hline
\end{tabular}
\label{tab1}
\end{table}

\begin{figure}
\centering
\includegraphics[width=0.5\textwidth]{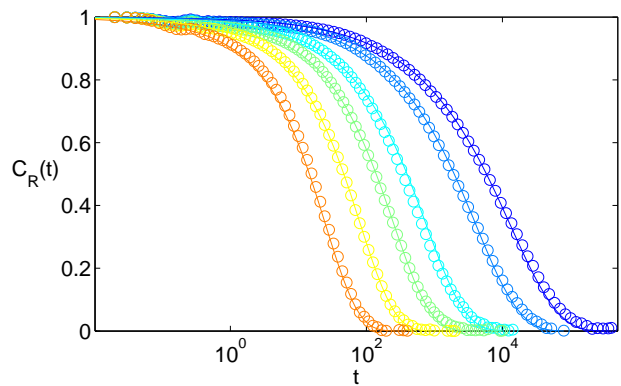}
\caption{(Color online). Comparison of the model relaxation function with data from the simulations}
\label{final}
\end{figure}

{\bf Summing up the three contributions}. Having modeled the spin de-correlation formulae
due to the two channels of decay for the three natural groups of quasi-particles, we should be able to predict the total relaxation function simply by summing up the three contributions, with each one weighted by the predicted concentration of the appropriate group of quasi-species with a given $k$. In other words, we should plot
\begin{equation}
C_R(t) = \sum_{k=0}^2 c_k(T) F_k(t) \ ,
\end{equation}  
and compare the results with the data. This comparison is shown in Fig. \ref{final}. 

We re-iterate that {\bf all} the temperature dependence in these plots stems from the dependence of the typical decay time of each elementary process  and from the temperature dependent weights of the concentration of $k$ groups of quasi-species. The apparent change of $\beta$ in the stretched exponential fit to this data \cite{06ST} as a function of temperature stems solely from the change of importance of the various $k$ groups as a function of the temperature, and from the change of importance of rotation vs. translation (remember that the translational elementary contributions appear here always as pure exponentials). 

The bottom line of this analysis is two-fold; first, we demonstrate that it is possible to break apart relaxation processes in glass forming system to more elementary contributions, understanding much better the origin o the complex time dependence of relaxation functions. In this point we simply extend what was proposed in 
\cite{08HP,08LP}. Second, we stress the danger of straight numerical fits to relaxation functions; when these change their functional form with temperature, the reason should be sought in the existence of a composite process with various dynamical contributions, each of which may be quite simple. Whether the system is 'fragile' or 'strong' in the Angell parlance \cite{96EAN} may be in the eyes of the beholder, especially if the said beholder did not reveal the details of the physical phenomenon.

This work had been supported in part by the Israel Science Foundation, the Minerva Foundation, Munich Germany and the Tauber Fund for research in the Science of Complexity. 


\end{document}